\documentstyle[12pt]{article}
\newcommand{\sptwo}{1.4}

\newcommand{\doublespace}{\edef\baselinestretch{\sptwo}\Large\normalsize}

\begin{document}
\doublespace

\begin{center}
{\bf NEUTRINO-EXCHANGE INTERACTIONS IN 1-, 2-, and 3-DIMENSIONS\\
  $~$}\\
Ephraim Fischbach and Brian Woodahl\\
Physics Department, Purdue University, West Lafayette, IN  47907
\end{center}
\medskip
\begin{center}
{\bf ABSTRACT}
\end{center}

\begin{quote}
$~~~~$ We examine several recent calculations of the self-energy
of a neutron star arising from neutrino-exchange.  It is shown
that the results of Abada, {\it et al.} in 1+1 dimensions have
no bearing on a 3-dimensional neutron star, since the criticality
parameter $G_F N/R^2$ is always much smaller than unity in 1+1
dimensions.  The calculation of Kiers and Tytgat
in 3-dimensions is shown to disagree with the lowest order
2-body contribution, which is known exactly.  This discrepancy
raises the possibility that the description of a neutron star
as a continuous medium may be inappropriate when calculating
higher-order many-body effects.  We conclude that none of the
recent calculations contradict the earlier claims that the
neutrino-exchange contributions to the self-energy of a
neutron star are unphysically large, when calculated in the
standard model.  The implication of this result, that neutrinos
must have a non-zero mass, $m_\nu  > 0.4~{\rm eV}/c^2$, remains intact.
\end{quote}

\pagebreak

It has been argued that many-body forces arising from the exchange
of massless neutrinos can lead to an unphysically large energy density
in white dwarfs and neutron stars [1, 2].  Intuitively this comes about
because in the absence of any other physical parameters, the $k$-body $(k = 2,\ 4,\ 6,\ ...)$ contribution $W^{(k)}$ to
the binding energy $W$ of a neutron star of radius $R$ must be of the form [1, 2]

$$  W^{(k)} ~~\propto~~ \frac{G_F^k}{R^{2k+1}} {N \choose k}.
    \eqno{(1)}
$$

\noindent
Here $G_F$ is the Fermi constant and ${N \choose k}$ is
the binomial coefficient,

$$  {N \choose k} =
     \frac{N!}{k!(N-k)!} \cong \frac{N^k}{k!} ~~~~\mbox{for}~~~
       N \gg k.
  \eqno{(2)}
$$

\noindent
Combining Eqs. (1) and (2) we find [1, 2]

$$  W^{(k)} \sim \frac{1}{k!} \frac{1}{R}
    \left( \frac{G_FN}{R^2}\right)^k.
    \eqno{(3)}
$$
For a typical neutron star $(G_FN/R^2) = {\cal O}(10^{13})$, from which it
follows that for $k \ll N$ higher-order many-body interactions
make increasingly larger contributions to $W^{(k)}$ [3]. The energy, $W = \Sigma_k W^{(k)}$ can exceed the mass of the neutron star,
as is shown explicitly in Ref. [1], 
and this eventually leads to the conclusion that neutrinos must
have a minimum mass, $m_\nu > 0.4~{\rm eV}/c^2$.
The effect of a nonzero $m_\nu$ is to produce a ``saturation"
of the neutrino-exchange force, and ultimately a physically
acceptable value for $W$.
%
%\stackrel{>}{\sim}
%

Following the publication of Ref. [1], a number of papers appeared
dealing with various questions raised by this calculation [4-10].
Here we focus on attempts to calculate the energy density $w$
non-perturbatively starting from variants of the Schwinger
formula [11] used in Ref. [1]:

$$  W = \frac{i}{2\pi} {\rm Tr} \left\{ \int_{-\infty}^{\infty} dE \  \ell n
    [ 1 + \frac{G_Fa_n}{\sqrt{2}} N_\mu \gamma_\mu
    (1+\gamma_5) S_F^{(0)} (E) ] \right\}.
    \eqno{(4)}
$$
In Eq. (4) $a_n = -1/2$ is the neutrino-neutron coupling constant, $N_\mu$
is the neutrino current, and $S_F^{(0)}(E)$ is the free neutrino
propagator.  In the notation of Ref. [1],
$N_\mu = i\rho(\vec{x})\delta_{\mu 4}$, where $\rho(\vec{x})$
is the neutron density, with $\int d^3x \rho(\vec{x})
= N = {\cal O}(10^{57})$.  The perturbative method of Ref. [1]
treats neutrons as discrete particles, and  $W$ can be calculated
for {\it any} $\rho(\vec{x})$. Our results, to be
presented elsewhere show that the result in Ref. [1] is
robust:  For a variety of symmetric and asymmetric matter
distributions the calculation of $W$ given in Ref. [1]
leads to an unphysically large value for massless
neutrinos.  

By constrast the non-perturbative calculations treat the neutron
star as a continuous medium whose density is constant, either
throughout all space [4, 8] or in some sub-region [7, 10].  Since
various cancellations occur in both approaches, one explanation
for the differences among the results may be the different
pictures used to describe a neutron star (i.e. discrete neutrons
versus a continuous medium), as we discuss in more detail
elsewhere.  Along with other considerations, treating the neutron
star as an object of finite extent is essential.

Two recent papers have attempted to calculate $W$ for a ``neutron
star" of finite extent, rather than calculating the energy
density of an infinite medium.  Abada, {\it et al.} [7] have calculated
the effect of many-body neutrino exchange in a 1+1 dimensional
``neutron star", and Kiers and Tytgat [10] have carried out a
similar calculation in 3+1 dimensions.

Returning to Eqs. (1)-(3) we note from the heuristic derivation of
Eq. (3) --- or from the formal treatment in Ref. [1] --- that
$W^{(k)}$ will have the same functional form in any
number of spatial dimensions.  The differences among the
1-, 2-, and 3-dimensional cases arise when $(G_FN/R^2)$ is
re-expressed in terms of the corresponding linear, surface and
volume densities:

\begin{quote}
\underline{1-dimension}: For a linear array of length $L$ and
density $\lambda$, $N=\lambda L$ and $R = L/2$ so that
\end{quote}
$$  G_F N/R^2 = 4G_F\lambda/L~.
   \eqno{(5)}
$$
The maximum value of $\lambda$ corresponds to taking the
inter-neutron spacing to be given by the hard core radius
$r_c = 0.5 \times 10^{-13}\ {\rm cm}$, which gives
$\lambda = 2 \times 10^{13} \ {\rm cm}^{-1}$.
Taking $R \equiv R_{10} = 10 \ {\rm km}$ then gives

$$  G_F N/R_{10}^2\vert_{1-{\rm dim}} = 2\times 10^{-25}~.
    \eqno{(6)}
$$
Hence in 1-dimension the ``criticality parameter"
$G_FN/R^2$ in Eq. (3) is always much less than unity,
from which one could conclude at the outset that $W$
would always remain acceptably small.  Physically this comes
about because in 1-dimension there are too few neutrons
in the neighborhood of a given neutron to allow the combinatoric
factor ${N \choose k}$ to provide
a sufficiently large enhancement.

From Eq. (6) we recognize that many-body ($k > 2$) effects are supressed in 1-dimension, and hence the leading contribution to the energy comes from the 2-body neutron-neutron potential [1, 2, 12-16],
$$	V^{(2)}(r) = {G_{F}^2 a_{n}^2 \over 4 \pi^3 r^5}.
	\eqno{(7)}
$$
The average interaction energy $U_{\rm 1-dim}^{(2)}$ of a pair of neutrons in a linear array of length $L$ would be given by

$$  U_{\rm 1-dim}^{(2)} = \int_{r_c}^{L} dr P_{1}(r) {G_{F}^2 a_{n}^2 \over 4 \pi^3 r^5}.
	\eqno{(8)}
$$
Where $P_{1}(r)$ is the 1-dimensional probability density given by [1]
$$	P_{1}(r) = {2 \over L} - {2 r \over L^2}.
	\eqno{(9)}
$$
The leading term to the average energy is found by combining Eqs. (8) and (9),
$$  U_{\rm 1-dim}^{(2)} = {G_{F}^2 a_{n}^2 \over 8 \pi^3 r_{c}^4 L}.
	\eqno{(10)}
$$
We note that this result is unaffected by inclusion of any contributions from a trapped neutrino sea, since the 2-body potential is dominated by small values of $r$ [9]. For a linear array of $N$ discrete neutrons, there are approximately $N^2/2$ pairs that can be formed, hence the energy in 1-dimension is
$$	W_{\rm 1-dim} = W_{\rm 1-dim}^{(2)} = {G_{F}^2 a_{n}^2 \lambda^2 L \over 16 \pi^3 r_{c}^4}.
	\eqno{(11)}
$$
Consequently, the energy density $w_{\rm 1-dim}$ is a constant, given by
$$	w_{\rm 1-dim} = {G_{F}^2 a_{n}^2 \lambda^2 \over 16 \pi^3 r_{c}^4}.
	\eqno{(12)}
$$
This result conflicts with the 1-dimensional results of Abada {\it et al.} [7] who find $w_{\rm 1-dim}$ to be identically zero. We note that since the Abada {\it et al.} results can  be expanded in a power series in $G_{F}$, they must agree with the lowest-order perturbative results given in Eqs. (11) and (12).

\begin{quote}
\underline{2-dimensions}: For a planar circular array of neutrons
with surface density $\sigma$, we have $N = \sigma \pi R^2$
so that
\end{quote}
$$  G_FN/R^2 = \pi G_F \sigma = \mbox{constant} .
    \eqno{(13)}
$$
Assuming, as before, a hard core radius $r_c = 0.5 \times 10^{-13} \ {\rm cm}$
gives $\sigma = 1 \times 10^{26}\ {\rm cm}^{-2}$, and for
$R = R_{10} = 10\ {\rm km}$ we find

$$   G_F N/R_{10}^2\vert_{2-{\rm dim}} = 2 \times 10^{-6}.
   \eqno{(14)}
$$
Since this is again smaller than unity, we conclude that
the neutrino-exchange energy $W$ is well-behaved in
2-dimensions and dominated by the 2-body contribution. The energy density in 2-dimensions $w_{\rm 2-dim}$,  can be obtained in a similar manner to the 1-dimensional result of Eq. (12), and we find
$$	w_{\rm 2-dim} = {G_{F}^2 a_{n}^2 \sigma^2 \over 12 \pi^2 r_{c}^3}.
	\eqno{(15)}
$$

\begin{quote}
\underline{3-dimensions}: For a spherical volume of radius $R$
and number density $\rho$ we have in 3-dimensions
$N = \frac{4}{3} \pi R^3 \rho$, which gives
\end{quote}
$$  G_F N/R^2 = (4/3) \pi G_F\rho R.
   \eqno{(16)}
$$
Using $\rho = 4 \times 10^{38}\ {\rm cm}^{-3}$ from Ref. [1] gives

$$ G_F N/R_{10}^2\vert_{3-{\rm dim}} = 8 \times 10^{12},
   \eqno{(17)}
$$
and hence it is only in 3 spatial dimensions that the possibility
exists for an unphysically large energy density arising from
the exchange of massless neutrinos.  Thus the recent
calculation of Abada, {\it et al.}, which is done in 1-dimension, has little bearing on the calculation in Ref. [1].

\bigskip
We turn next to the paper of Kiers and Tytgat (KT) [10] which
also calculates $W$ under the assumption that a neutron
star can be described as a continuous medium. KT work in a finite 3-dimensional volume, and hence their starting point
is closest to that of Ref. [1] amongst the recent papers which
use the continuous medium approximation.  Since KT do not find
a large value for $W$, in contrast to Ref. [1], it is instructive
to compare the discrete and continuum calculations of $W$.

For present purposes we focus on the 2-body contribution $W^{(2)}$ to the self-energy of a neutron star $W$, which can be calculated {\it exactly} with no approximations [2].  The 2-body potential is given by Eq. (7). The average interaction energy $U^{(2)}$ of a pair of neutrons having a uniform density distribution
in a spherical volume of radius $R$ is then given by
[1, 2]

$$  U^{(2)} = \int_{r_c}^{2R} dr P(r) \frac{(G_Fa_n)^2}{4\pi^3r^5},
   \eqno{(18)}
$$
where $P(r)$ is the probability density function\footnote{The normalization factor $\eta(r_{c}, R)$ was  approximated by unity in Refs. [1] and [2].} given by [1, 2]

$$  P(r) = \frac{3r^2}{R^3} \left[ 1-\frac{3}{2}
    \left( \frac{r}{2R}\right) + \frac{1}{2}
    \left( \frac{r}{2R}\right) ^3 \right] \eta(r_{c}, R),
   \eqno{(19{\rm a})}
$$
$$  \eta(r_{c}, R) = {1 \over 1-8 \left( \frac{r_c}{2R}\right)^3+9 \left( \frac{r_c}{2R}\right)^4-2 \left( \frac{r_c}{2R}\right)^6}.
    \eqno{(19{\rm b})}
$$
Combining Eqs.(18) and (19) we find that the energy per pair
of neutrons, $U^{(2)}$ is {\it exactly} given by
$$  U^{(2)} = \frac{3}{8\pi^3} \frac{(G_Fa_n)^2}{\hbar c}
     \frac{1}{R^3 r_c^2} 
     \left(1- \frac{r_c}{2R}\right)^3 \eta(r_{c}, R),
    \eqno{(20)}
$$
where we have reinstated the factor $\hbar c$.
For a neutron star containing $N$ discrete neutrons, there are
$N(N-1)/2$ pairs and hence the 2-body contribution
$W^{(2)}$ to $W$ is given by

$$  W^{(2)} = \frac{3}{16\pi^3}
              \frac{(G_Fa_n)^2}{\hbar c}
              \frac{N(N-1)}{R^3r_c^2}  \left( 1- \frac{r_c}{2R}\right)^3
		\eta(r_{c}, R),
      \eqno{(21)}
$$
which is also an {\it exact} result.  Since $N \gg 1$
for neutron star, we can replace $N(N-1)$ by $N^2$, and
re-express $N$ in terms of the number density $\rho$:

$$  W^{(2)} = \frac{1}{3\pi} \frac{(G_Fa_n\rho)^2}{\hbar c}
              \frac{R^3}{r_c^2} \left(1- \frac{r_c}{2R}\right)^3
		\eta(r_{c}, R).
    \eqno{(22)}
$$
Also of interest is the energy density $w^{(2)} = W^{(2)}/V$,

$$  w^{(2)} = \frac{1}{4\pi^2}
    \frac{(G_Fa_n\rho)^2}{\hbar c\,r_c^2} \left( 1-\frac{r_c}{2R}\right)^3
	\eta(r_{c}, R).
   \eqno{(23)}
$$

\noindent
It follows from Eq. (23) that the energy density $w^{(2)}$ is an increasing function of $R$, and approaches a constant as $R \rightarrow \infty$.

We now compare these results to those obtained by KT for the
2-body contribution.  KT find for $W^{(2)}$

$$ 
    W^{(2)}\vert_{\rm KT} \approx 
    \biggl({G_F\rho a_n \over \sqrt{2}}\biggr)^2 R^2 \Lambda,
   \eqno{(24)}
$$
where $\Lambda$ is a cutoff which plays a role similar to $r_c$
in Eq. (21).  KT then use Eq. (24) to conclude that

$$
     w^{(2)}\vert_{\rm KT} \approx
     \biggl({G_F\rho a_n \over \sqrt{2}}\biggr)^2 \frac{\Lambda}{R}
     \stackrel{R \rightarrow \infty}{\longrightarrow} 0 .
   \eqno{(25)}
$$

\noindent
Comparing Eqs.(24) and (25) to Eqs.(22) and (23), respectively,
we see that the $R$-dependence of the KT results disagrees
with that of the exact results obtained by assuming discrete neutrons. Since the KT results can be expanded in a power series in $G_{F}$, the lowest order KT results must agree with the corresponding perturbative results given in Eqs. (22) and (23). Furthermore the KT results are seen to {\it underestimate}
$W^{(2)}$ and the energy density $w^{(2)}$ as $R$ increases.
This is a particularly troublesome point since it raises the
possibility that the many-body contributions may be similarly
underestimated in the KT formalism.

It should be emphasized that the discrepancy between the KT
results and the exact results is a matter of principle,
and cannot be dismissed on the grounds that $W^{(2)}$ is
in either case small.  Until the origin of this discrepancy
is fully understood one cannot be certain that it does not
affect other parts of their calculation as well.

\medskip
\begin{center}
{\bf CONCLUSIONS}
\end{center}

The recent papers by Abada, {\it et al.} [7], and by
Kiers and Tytgat [10] have been analyzed. We have shown explicitly that the results of Ref. [7] in 1-dimension have no bearing
on whether the neutrino-exchange energy density in 3-dimensions
is unphysically large in a neutron star.  The KT calculation,
although carried out in 3-dimensions, fails to reproduce the
2-body result which is known exactly and, moreover, underestimates
this contribution.  Obtaining the correct 2-body result is a
litmus test for the validity of any calculational scheme,
and hence the significance of the KT calculation is unclear at present.
One question that must be explored is the validity of approximating
the neutron star by a continuous medium.  Although such a
picture is routinely employed in discussing the MSW effect
(in which $G_F$ enters only in lowest order), its appropriateness
for calculating self-energy interactions involving higher-order weak processes remains to be demonstrated.
Another question that remains to be understood is where the
medium description breaks down:  Clearly an atomic nucleus should
be described in terms of discrete neutrons
for purposes of calculating $W$.  Since a neutron star
behaves in some way as a big nucleus, the transition from
a description in terms of discrete neutrons to one in terms of
a medium must be fully understood.
These and other related questions will be explored in more 
detail elsewhere.

\pagebreak

\begin{center}
{\bf REFERENCES}
\end{center}

\begin{enumerate}
%1
\item E. Fischbach, Ann. Phys. {\bf 247}, 213 (1996).
%2
\item E. Fischbach, D.E. Krause, C. Talmadge, and
D. Tadi\'c, Phys. Rev. {\bf D52}, 5417 (1995).
%3
\item R.P. Feynman, F.B. Morinigo, and W.G. Wagner,
``Feynman Lectures on Gravitation" (Addison-Wesley,
Reading, MA, 1995).
%4
\item As. Abada, M.B. Gavela, and O. P\`ene, Phys. Lett.
{\bf B387}, 315 (1996).
%5
\item J.A. Grifols, E. Mass\'o, and R. Toldr\`a,
Phys. Lett. {\bf B389}, 563 (1996).
%6
\item A. Smirnov and F. Vissani, preprint hep-ph/9604443 v.2.
%7
\item A.S. Abada, O. P\`ene, and J. Rodriguez-Quintero,
preprint hep-ph/9712266.
%8
\item M. Kachelriess, preprint hep-ph/9712363.
%9
\item B. Woodahl, M. Parry. S.-J. Tu, and E. Fischbach,
preprint hep-ph/9709334.
%10
\item K. Kiers and M. Tytgat, preprint hep-ph/9712463.
%11
\item J. Schwinger, Phys. Rev. {\bf 94}, 1362 (1954).
%12
\item G. Feinberg and J. Sucher, Phys. Rev. {\bf 166},
1638 (1968).
%13
\item G. Feinberg, J. Sucher, and C.-K. Au,
Phys. Rep. {\bf 180}, 83 (1989).
%14
\item S.D.H. Hsu and P. Sikivie, Phys. Rev. {\bf D49},
4951 (1994).
%15
\item J.B. Hartle, Phys. Rev. {\bf D1}, 394 (1970).
%16
\item C.J. Horowitz and J. Pantaleone, Phys. Lett.
{\bf B319}, 186 (1993).
\end{enumerate}

\end{document}